\documentstyle [12pt,a4] {article}

\renewcommand{\(}{\left(}
\renewcommand{\)}{\right)}
\newcommand{\g}{{\bf g}}

\newcommand{\su}{{\bf su}}
\newcommand{\ox}{\otimes}

\newcommand{\qqqquad}{\quad\quad\quad\quad}
\newcommand{\ad}{\mbox{ad}}

\def\dj{d\kern-.30em\raise1.25ex\vbox{\hrule width .3em height .03em}}
\def\Dj{D\kern-.75em\raise0.75ex\vbox{\hrule width .3em height .03em}\kern.60em}

\begin{document}
\title{\begin{flushright}\small hep-th/9601033 \end{flushright}
  \smallskip
{\bf SU$_q(n)$ Gauge Theory}}
\author {Anthony Sudbery\\[10pt] 
  \small Department of Mathematics\\[-2pt] \small University of York\\[-2pt]
  \small Heslington\\[-2pt] \small York\\[-2pt] \small England\\[-2pt]
  \small YO1 5DD\\ 
  \small  E-mail:  as2@york.ac.uk}
\date{29 February 1996}
\maketitle
\bigskip
\begin{abstract}
A field theory with local transformations belonging to the quantum group
SU$_q(n)$ is defined on a classical spacetime, with gauge potentials
belonging to a quantum Lie algebra. Gauge transformations are defined for
the potentials which lead to the appropriate quantum-group
transformations for field strengths and covariant derivatives, defined for all elements of
SU$_q(n)$ by means of the adjoint action. This guarantees a non-trivial
deformation. Gauge-invariant commutation relations are identified.
\end{abstract}
\bigskip

\section{Introduction}

A number of authors [1--11, 13, 14] have proposed $q$-deformations of gauge 
field theory. One motivation for this is the prospect of a theory with 
as large a set of invariances as conventional gauge theory---and therefore, 
hopefully, the same renormalisability properties---but in which the symmetry 
of physical predictions is broken without the intervention of extra fields 
such as the Higgs field. Another motivation is the hope that a consistent 
quantum theory of gravity may take such a form. The most successful of
these theories \cite{BrzezMaj:quant, Durdevic, Hajac} have been those in which the
structure of space-time is made non-commutative. This is an interesting
feature if one's aim is to construct a theory of quantum gravity; but
for the first motivation mentioned above, it would be more natural to
construct the theory on a classical space-time. All attempts so far 
to do this have failed. In some cases \cite{Durdevic, Watamura:BRST} the theory does not
have a good classical limit as $q \rightarrow 1$. This is not surprising
for a theory based on a bicovariant calculus on a quantum group, as such
calculi do not normally have a good classical limit; only for the
quantum groups GL$_q(n)$ does one obtain a Lie algebra with the same
dimension as the classical one, and therefore a deformed gauge theory
with the same number of gauge fields as the classical theory. Other
theories \cite{Hirayama, IsaevPop1, IsaevPop2} have the opposite problem: 
the classical limit exists, but the
deformed theory never gets away from it, remaining always isomorphic to
it. The reason for this, as was explained by Brzezinski and Majid
\cite{BrzezMaj:class}, is that such theories use a notion of gauge
transformation modelled too closely on the classical form of
conjugation, namely $F \mapsto gFg^{-1}$ where $g$ is an invertible element
of the quantum group, instead of the adjoint action $F \mapsto
g_{(1)}FS(g_{(2)})$ which is more appropriate for a quantum group with
its Hopf algebra structure (and, indeed, for a classical group when one 
wants to consider non-invertible gauge transformations such as 
infinitesimal ones). This adjoint type of gauge transformation has been
thought to be impossible because "there is no way to define $F$ from
[the gauge field] to transform as desired". In this paper a way is found.

The basis of the theory presented here is a new notion of quantum Lie algebra
\cite{qliealg} which does not require the existence of a bicovariant 
differential calculus. This leads to a very natural definition of a gauge 
transformation for gauge potentials in a classical space-time
such that the transformation at each point is an element of a quantum group instead 
of a group, and hence to a $q$-deformed gauge theory based entirely on 
the Hopf algebra structure of a quantum groups. The theory contains
gauge potentials whose gauge transformations imply the appropriate Hopf
adjoint-type transformation properties for the field strengths. The required quantum Lie 
algebra is known to exist for the quantum groups SU$_q(n)$, and is conjectured 
to exist for the $q$-deformations of all simple Lie groups.

The geometrical picture to bear in mind in considering this theory is
similar to the geometry of fibre bundles 
underlying classical gauge theory; it is a picture of vector bundles over a 
classical manifold, the structure group by which the vectors transform being
replaced by a Hopf algebra (generalising the group algebra of the structure 
group or---what is conceptually almost the same thing---the enveloping algebra 
of the Lie algebra of the structure group). If this Hopf algebra is the 
quantised enveloping algebra U$_q(\su(n))$ it can also be regarded as 
an enveloping algebra, being generated by a quantum Lie algebra \cite{qliealg}; 
this makes it possible to define a connection in the Hopf vector bundle 
with a similar geometrical meaning to the connection in a vector bundle 
with a classical structure group. This geometry is classical in the sense 
that it has points in the base manifold; it is not necessary, as it usually 
is in non-commutative geometry, to renounce points in favour of functions.

The resulting gauge theory can only be considered as a quantum field theory; 
it cannot be defined as a classical field theory because the components of 
the fields cannot be taken to commute. For consistency with the gauge transformations, 
they must satisfy $q$-commutation relations. In the case of the 
gauge potentials, it is not even possible to impose the full set of $q$-commutation 
relations; roughly speaking, half of the commutators must be unspecified. 
This feature was also found by Castellani \cite{Castellani:UqN} in his
$U_q(n)$ gauge theory. Unlike Castellani, we do not need to suppose that 
the parameters of our gauge transformations are non-commuting objects.

The theory has the desired feature of a set of gauge transformations as 
large as in classical gauge theory, while exhibiting symmetry breaking 
in its physical predictions. However, this symmetry breaking is not sufficient 
to provide a non-zero mass for the gauge fields without some further mechanism 
such as a Higgs field.

\section{$q$-Gauge Transformations}

Let ${\cal H} = U_q(\g )$ be the quantised enveloping algebra of the 
$N$-dimensional Lie algebra $\g$. This is a Hopf algebra, with a coproduct 
$\Delta x = \sum x_{(1)}\ox x_{(2)}$ and an antipode $S(x)$, and it acts 
on itself by the adjoint action
\begin{equation}
\ad x(y) = x_{(1)}yS(x_{(2)})
\end{equation}
where, as is customary, we have omitted the $\sum$ sign in the sum over 
the terms in the coproduct on the right-hand side. A {\em quantum Lie algebra} 
$\g_q$ is an $N$-dimensional subspace of $\cal H$ with the following 
properties:
\begin{enumerate}
  \item $\g_q$ is invariant under the adjoint action, so that the quantum 
Lie bracket
    \begin{equation}\label{qbrac}
        [x,y]_q=\ad x(y) \in \g_q
    \end{equation}
  is defined for all $x,y \in \g_q$.
  \item $\g_q$ generates ${\cal H}$ as an algebra.
  \item The elements of $\g_q$ satisfy relations of the form 
\begin{equation}\label{quom}
  X_i X_j - \sigma _{ij}^{kl}X_k X_l = C[X_i, X_j]
\end{equation}
where $C$ is a central element of ${\cal H}$ (a function of the Casimirs 
of $\g_q$), the $X_i$ are basis elements of $\g_q$, and $\sigma _{ij}^{kl}$ 
is an $N^2 \times N^2$ matrix which has $1$ as an eigenvalue (a deformation 
of the classical flip operator: $\sigma _{ij}^{kl}=\delta _i^l \delta 
_j^k$ when $q=1$).
\item The quantum Lie bracket (\ref{qbrac}) is antisymmetric with respect 
to $\sigma$ in the sense that 
\begin{equation}\label{antisym}
t^{ij}[X_i,X_j]_q=0\qqqquad\mbox{whenever}\qquad t^{ij}\sigma _{ij}^{kl}=t^{kl}.
\end{equation}
\end{enumerate}

It is known \cite{qliealg} that a quantum Lie algebra $\g_q$ exists 
for $\g = \su (n)$ (though the sense in which it generates $U_q(\g)$
has yet to be rigorously defined) and that in this case the coproducts of the elements 
of $\g_q$ are of the form
\begin{equation}\label{cop}
\Delta (X_i) = X_i \ox C + u_i^j \ox X_j .
\end{equation}
The quantum flip operator $\sigma $ is related to the adjoint action of 
the elements $u_i^j$:
\begin{equation}\label{flip}
\ad u_i^j(X_k) = \sigma _{ik}^{lj}X_l.
\end{equation}
We will illustrate the constructions in this paper by the case of $\g 
= \su (2)$, for which the quantum Lie algebra $\su (2)_q$ has a 
basis $\{X_0, X_+, X_-\}$ with brackets 
\begin{equation}\label{sl_q(2)}
  \begin{array}{ccc}
    {[}X_+,X_+]=0,     &  [X_+,X_0]=-q^{-1}X_+,   & 
      [X_+,X_-]=(q+q^{-1})X_0, \\[5pt]
    {[}X_0,X_+]=qX_+,& [X_0,X_0]=(q-q^{-1})X_0,& 
      [X_0,X_-]=-q^{-1}X_-, \\[5pt]
    {[}X_-,X_+]=-(q+q^{-1})X_0, & \:[X_-,X_0]=qX_-, &
      [X_-, X_-]=0,
  \end{array}
\end{equation}
central element $C$ given by
\begin{equation}
C^2 = 1 + (q-q^{-1})^2\(X_0^2 + \frac{qX_-X_+ + q^{-1}X_+X_-}{q+q^{-1}}\),
\end{equation}
and relations
\begin{eqnarray}\label{quom2}
  qX_0X_+-q^{-1}X_+X_0&=&CX_+,\nonumber\\[3pt]
  q^{-1}X_0X_--qX_-X_0&=& - CX_-,\\[3pt]
  X_+X_--X_-X_++(q^2-q^{-2})X_0^2&=&(q+q^{-1})CX_0.\nonumber
\end{eqnarray}

We would like to make the quantum group ${\cal H}=U_q(\g)$ into a gauge group by 
taking gauge transformations to be elements $h(x)$ of $\cal H$ depending on 
the space-time point $x$, i.e. polynomials (or 
power series) in the Lie algebra elements $X_i$ with $x$-dependent coefficients. 
A gauge potential $A_\mu (x)$ should be a space-time vector field with 
values in the quantum Lie algebra $\g_q$:
\begin{equation}
A_\mu (x) = A_\mu ^i(x)X_i
\end{equation}
where the $A_\mu^i(x)$ are ordinary fields. It should transform 
under the gauge transformation $h(x)$ by
\begin{equation}\label{gtrans}
A_\mu (x) \mapsto A^\prime _\mu (x) = h(x)_{(1)}A_\mu (x) S\(h(x)_{(2)}\)
          - \alpha ^{-1}S(C)^{-1}\partial _\mu \(h(x)_{(1)}\)S\(h(x)_{(2)}\)
\end{equation}
where $\alpha $ is a coupling constant. However, the second term here 
is not well-defined, since we do not know how to distribute the $x$-dependence 
of the coproduct $\Delta (h(x))$ between the two factors before differentiating 
the first factor. Moreover, the set of gauge transformations $h(x)$ with 
values in $\cal H$ would be very much bigger than the classical gauge group 
of functions $g(x)$ with values in the group $G$ whose Lie algebra is 
$\g$. Such functions can be written as $g(x)=$ exp$X(x)$ where $X(x)$ is 
a function with values in the Lie algebra $\g$. We therefore restrict 
ourselves to gauge transformations of the form 
\begin{equation}\label{h(x)}
h(x)=f(X(x))
\end{equation}
where $f$ is a polynomial or power series function and $X(x)$ is a function 
on space-time with values in the quantum Lie algebra $\g_q$:
\begin{equation}
  X(x)=\xi ^i(x)X_i.
\end{equation}
We define the coproducts of such infinitesimal gauge transformations by
\begin{equation}
\Delta (X(x)) = \xi ^i(x)X_i \ox C + u_i^j \ox \xi^i(x)X_j
\end{equation}
(see (\ref{cop})), and extend to powers of $X(x)$ multiplicatively. Then 
if $h(x)$ is as in (\ref{h(x)}) we can form 
\[
m \circ (\partial_\mu \ox S)\Delta h(x) =
 \sum \partial_\mu \(h(x)_{(1)}\)S\(h(x)_{(2)}\)
\]
where $m:\cal H \ox H \rightarrow H$ denotes multiplication in
${\cal H}$. (What we are doing is to  
regard $X$ as an element of ${\cal H}(\cdot )={\cal F \ox H}$ where 
${\cal F}$ is an algebra 
of c-number functions on space-time; then we have defined $\Delta X(\cdot )$ 
and $\Delta h(\cdot )$ as elements of ${\cal H}(\cdot ) \ox {\cal H}(\cdot )$ and we can apply 
$\partial_\mu \ox S$ to $\Delta  h(\cdot )$ before multiplying the two factors.) 
Define
\[
\delta_\mu h(x) = S(C)^{-1}m\circ (\partial_\mu \ox S)\Delta h(x).
\]
We will show that $\delta_\mu h(x)$ belongs to the quantum Lie algebra 
$\g_q$.

We consider powers of infinitesimal gauge transformations, $h_n(x)=X(x)^n$, 
and proceed by induction on $n$. For $n=1$ we have 
\[
h_1(x)=\xi^i(x)X_i,
\]
\begin{eqnarray*}
\delta_\mu h_1(x) &=& S(C)^{-1}m\circ (\partial_\mu\ox S)\(\xi^i(x)\ox C 
+ u_i^j\ox\xi^i(x)X_j\)\\
&=& S(C)^{-1}\partial_\mu\xi^iX_i S(C).
\end{eqnarray*}
But $S(C)$ is central if $C$ is, for 
\begin{eqnarray*}
S(C)x &=& x_{(1)}S(C)S(x_{(2)})x_{(3)} = x_{(1)}S(x_{(2)}C)x_{(3)} = x_{(1)}S(Cx_{(2)})x_{(3)}\\
&=& x_{(1)}S(x_{(2)})S(C)x_{(3)} = \varepsilon (x_{(1)})S(C)x_{(2)} = 
S(C)x.
\end{eqnarray*}
Hence
\begin{equation}
\delta_\mu h_1(x) = \partial_\mu\xi^i(x)X_i \in \g_q.
\end{equation}

Now suppose $\delta_\mu h_m(x)$ and $\delta_\mu h_n(x)$ both belong 
to $\g_q$, and consider the product $h_{m+n}(x)=h_m(x)h_n(x)$. We have
\begin{eqnarray*}
  \delta_\mu h_{m+n}(x) &=& S(C)^{-1}m\circ (\partial_\mu \ox S)
   \(h_m(x)_{(1)}h_n(x)_{(1)} \ox h_m(x)_{(2)}h_n(x)_{(2)}\)\\
  &=& S(C)^{-1}\(\partial_\mu h_m(x)_{(1)}h_n(x)_{(1)} + 
    h_m(x)_{(1)}\partial_\mu h_n(x)_{(1)}\)\\
  && S\(h_n(x)_{(2)}\)S\(h_m(x)_{(2)}\)\\
  &=& \varepsilon (h_n(x))\delta_\mu h_m(x)
    + \ad h_m(x).\delta_\mu h_n(x)
\end{eqnarray*}
which belongs to
 $\g_q$ since $\g_q$ is ad-invariant.

It follows that the transformation law (\ref{gtrans}) is well-defined 
for gauge transformations of the type (\ref{h(x)}), i.e. that $A^\prime_\mu
(x)$ is a gauge potential (belongs to $\g_q$) if $A_\mu(x)$ is.
\section {Covariant Derivatives and Field Strengths}

Let $\Psi$ be a multiplet of fields transforming according to a representation 
$\rho$ of the Hopf algebra $\cal H$. Associated with this representation are 
the scalars
\[
c_\rho =\rho (C), \qqqquad \qqqquad \widetilde{c}_\rho =\rho (S(C)).
\]
We define the covariant derivative
\begin{equation}
D_\mu \Psi = \partial_\mu \Psi + \alpha \widetilde{c}_\rho \rho (A_\mu)\Psi.
\end{equation}
Assume that a product of fields transforms under a gauge transformation 
$h(x)$ according to the coproduct in $\cal H$: if fields $\phi (x), \psi (x)$ 
transform by 
\[
  \phi (x) \mapsto T_\phi [h(x), \phi (x)], \qqqquad 
  \psi (x) \mapsto T_\psi [h(x), \psi (x)]
\]
then the product $\phi (x)\psi (x)$ transforms by
\begin{equation}\label{product}
\phi (x) \psi (x) \mapsto T_\phi [h(x)_{(1)}, \phi(x)]T_\psi [h(x)_{(2)}, 
\psi (x)].
\end{equation}
Then the covariant derivative transforms by 
\begin{eqnarray}
  D_\mu \Psi (x) &\mapsto& \partial_\mu \big[\rho (h(x))\Psi (x)\big] 
    + \alpha \rho \(T_A[h(x)_{(1)},A_\mu (x)]\) T_{\Psi}[h(x)_{(2)}, \Psi (x)] \nonumber 
\\ 
  &=& \partial_\mu\rho (h(x))\Psi (x) 
    + \rho (h(x))\partial_\mu\Psi (x)  \nonumber\\
  && {} + \alpha \widetilde{c}_\rho \rho \(h_{(1)}A_\mu S(h_{(2)})   
     - \alpha ^{-1}S(C)^{-1}\partial_\mu h_{(1)}) S(h_{(2)})\)
       \rho \(h_{(3)}\)\Psi  \nonumber \\
  &=& \rho (h(x))D_\mu \Psi (x).
\end{eqnarray}
Thus $D_\mu \Psi$ is a covariant derivative in the sense that it transforms 
in the same way as $\Psi$.

To define the field strengths, it is necessary to introduce a second bracket 
$\{,\}_q$ on the quantum Lie algebra $\g_q$ by
\begin{equation}\label{newbrac}
\{X_i,X_j\}_q = \tau _{ij}^{kl}[X_k, X_l]_q
\end{equation}
where $\tau $ is an $N^2 \times N^2$ matrix satisfying
\begin{equation}\label{tau}
  (\sigma -1)(1-c_0\tau +\tau \sigma ) = 0,
\end{equation}
$c_0$ being the value of the central element $C$ in the 
$N$-dimensional adjoint representation ($c_0=q^2-1+q^{-2}$ for 
$\g=\su (n)$ \cite{qliealg}). This $\tau$ is 
arbitrary to the extent that one can add to it any solution $\tau ^\prime$ 
of 
\[
(\sigma -1)\tau ^\prime (\sigma - c_0) = 0
\]
but in general (i.e. if $c_0$ is not an eigenvalue of $\sigma $) this 
will imply $\sigma  \tau^\prime  = \tau^\prime $ and therefore by (\ref{antisym}) 
the bracket $\{,\}_q$ of (\ref{newbrac}) will be unchanged. In the case of $\su 
(2)_q$, when $\sigma$ has just the two eigenvalues $1$ and $-c_0$, eq. 
(\ref{tau}) is satisfied by $\tau =(2c_0)^{-1}$ and so in this case the 
new bracket is just a multiple of the quantum Lie bracket:
\begin{equation}
 \{X, Y\}_q = \frac{[X, Y]_q}{2c_0}.
\end{equation}

The field strengths are now defined by 
\begin{equation}\label{field}
 F_{\mu \nu } = \partial _\mu A_\nu - \partial _\nu A_\mu 
  + \alpha \Big(\{A_\mu , A_\nu \}_q - \{A_\nu , A_\mu \}_q\Big).
\end{equation}
To investigate their transformation properties, we need only consider 
an infinitesimal gauge transformation $h(x)=\xi ^i(x) X_i$ since the general 
gauge transformation is built up from these. For such $h(x)$ the transformation 
law (\ref{gtrans}) for the potentials becomes
\[
  A_\mu (x) \mapsto [h(x), A_\mu (x)]_q - \alpha ^{-1}\partial _\mu h(x)
\]
so that
\[
\{A_\mu (x), A_\nu (x)\}_q \mapsto \{\ad h_{(1)}(A_\mu ) - \alpha ^{-1}\delta_\mu h_{(1)},
  \; \ad h_{(2)}(A_\nu ) -\alpha^{-1}\delta_\nu h_{(2)}\}_q
\]
\[
= \ad h(x)\{A_\mu , A_\nu \}_q - \alpha ^{-1}\{\partial _\mu h(x), c_0 
A_\nu \}_q - \alpha ^{-1}\{\ad u_i^j(A_\mu ), \partial _\nu \xi ^i X_j 
\}_q
\]
using the coproducts (\ref{cop}) again, and the fact that for the form 
of $h(x)$ we are considering $h(x)_{(1)}$ and $h(x)_{(2)}$ are not both 
$x$-dependent in any term of the coproduct. Using eq. (\ref{flip}) for $\ad 
u_i^j$ and the definition (\ref{newbrac}--\ref{tau}) of the bracket $\{,\}_q$, 
we find that the field strengths transform as they ought to, like ordinary 
matter fields in the adjoint representation:
\begin{equation}\label{ftrans}
  F_{\mu \nu } \mapsto F_{\mu \nu }^\prime = \ad h(x) F_{\mu \nu }.
\end{equation}

It follows as usual that a gauge-invariant Lagrangian can be constructed
by forming a function of $F_{\mu \nu}$, $\Psi$ and $D_\mu \Psi$ which
is invariant under constant transformations by elements of $\cal H$.
Such a function of the field strengths alone, for example, is 
\begin{equation}
  {\cal L}_F = t_{ij}F_{\mu\nu}^iF^{j\mu\nu}
\end{equation}
where 
\[
  t_{ij} = \mbox{tr}(T \ad X_i \ad X_j ),
\]
$T=\sum S({\cal R}_{2}){\cal R}_{1}$ being the quantum trace element of
${\cal H}$ and ${\cal R}=\sum {\cal R}_1 \ox {\cal R}_2$ its universal $R$-matrix.
For ${\cal H}=U_q({\su (2)})$ this Lagrangian is 
\begin{equation}
  {\cal L}_F = (q+q^{-1})(qF^-.F^+ + q^{-1}F^+.F^-) + F_0^2.
\end{equation}

\section {Commutation Relations}

Doubts might be felt about our assumption of (\ref{product}) for the transformation 
properties of products of fields. In a quantum field theory, where symmetry 
transformations are implemented by unitary operators, there is no problem: 
if the gauge transformation $h(x)$ corresponds to an operator $U(h)$ and 
fields transform by
\begin{equation}
\psi (x) \mapsto U(h_{(1)})\psi (x)U(Sh_{(2)}),
\end{equation}
then the transformation law for products is a consequence of this. In 
a classical theory this is not so clear; moreover, the transformation 
law for products may not be consistent with the fact that the components 
of a classical field should commute. In a quantum field theory this can 
be allowed for by changing the commutation relations of the field components.

Let $\Psi_\alpha ^i(x)$ be a multiplet of fields transforming according to a representation 
$\rho $ of the Hopf algebra $\cal H$:
\begin{equation}\label{Psi}
U(h_{(1)})\Psi_\alpha ^i(x) U(Sh_{(2)}) = \Psi_\alpha ^j (x)\rho _j^i(h(x))
\end{equation}
$\alpha$ being a space-time (tensorial or spinorial) index. In the undeformed theory these fields 
will satisfy commutation relations of the form
\begin{equation}\label{CR}
[\Psi_\alpha ^i(x), \Psi_\beta ^j(y)] = \Delta _{\alpha \beta }^{ij}(x-y)
\end{equation}
if the fields $\Psi_\alpha ^i$ are bosonic (the fermionic case can be 
treated in a similar way). We can separate this into two equations, one 
symmetric and one antisymmetric in space-time indices:
\begin{equation}
\Big[\Psi_\alpha^i(x)\Psi_\beta^j(y)\pm\Psi_\beta^i(y)\Psi_\alpha^j(x)\Big]\Pi_{ij}^{\pm 
kl}
 =\(\Delta _{\alpha \beta }^{ij} \pm \Delta _{\beta \alpha }^{ij}\)\Pi_{ij}^{\pm 
kl}
\end{equation}
where $\Pi_{ij}^{\pm kl}=\delta _i^k\delta _j^l \pm \delta _i^l\delta 
_j^k$ are the projectors onto the symmetric and antisymmetric subspaces 
of the tensor product $V \ox V$ of the space $V$ carrying the representation 
$\rho $. These equations can easily be made compatible with the quantum 
group transformations; it is only necessary to replace $\Pi_{ij}^{\pm 
kl}$ by the projectors onto the corresponding invariant subspaces under 
the action of the quantum group (the $q$-symmetric and $q$-antisymmetric 
subspaces).

For the gauge potentials $A_\mu ^i$, the inhomogeneous term in the transformation 
(\ref{gtrans}) spoils this compatibility. In order to restore it for 
one of the sets of commutation relations, it is necessary to make the 
coupling constant $\alpha$ an operator with the commutation relation 
\begin{equation}
\alpha A_\mu ^i = c_0 A_\mu ^i\alpha .
\end{equation}
Then the effect of the transformation (\ref{gtrans}) on the commutator
\[
\left[ A_\mu ^i(x) A_\nu ^j(y) \mp A_\nu ^i(y) A_\mu ^j(x)\right]\Pi _{ij}^{\pm kl}
\]
is to produce a term quadratic in $A$ to which the same considerations 
apply as to a matter multiplet, the linear term
\begin{eqnarray*}
-\Big(A_\mu^m(x)[\ad h(x)_{(1)}]_m^i \alpha ^{-1}[\delta _\nu h(y)_{(2)}]^j
 + \alpha ^{-1}[\delta_\mu h(x)_{(1)}]^i A_\nu^m(y)[\ad h(y)_{(2)}]_m^j\\ 
 \mp (\mu  \leftrightarrow \nu , x \leftrightarrow y)\Big)\Pi _{ij}^{\pm kl}
\end{eqnarray*}
and an inhomogeneous term 
\[
\alpha ^{-2}\([\delta _\mu h(x)_{(1)}]^i[\delta _\nu h(y)_{(2)}]^j 
\mp (\mu \leftrightarrow \nu, x \leftrightarrow y )\)\Pi _{ij}^{\pm kl}.
\]
If we consider an infinitesimal gauge transformation $h(x)=\xi ^i(x) X_i$, 
the inhomogeneous term vanishes since $h_{(1)}$ and $h_{(2)}$ are not 
non-constant simultaneously, and the linear term becomes
\[
c_0 \alpha ^{-1}\left[ A_\mu ^m(x) \partial _\nu \xi ^n(y) - 
  A_\nu^m(y)\partial_\mu\xi^n(x)\right]
(\sigma _{lk}^{ij} \mp \delta _l^i \delta _k^j)\Pi _{ij}^{\pm kl}.
\]
For the adjoint representation ($V=\g_q$) the $q$-symmetric and $q$-antisymmetric 
subspaces of $V \ox V$ are respectively the kernel and the image of the 
quantum flip $\sigma$, so $(\sigma - 1)\Pi ^+ = 0$. Thus a commutation relation
\begin{equation}
\left[ A_\mu^i(x) A_\nu^j(y) - A_\nu ^i(y) A_\mu ^j(x)\right] \Pi _{ij}^{+ kl} = \mbox{c-number}
\end{equation}
is invariant under the gauge transformation (\ref{gtrans}). However, $\sigma 
$ does not have $-1$ as an eigenvalue if $q \ne 1$, and so the commutation 
relation
\begin{equation}
\left[ A_\mu ^i(x) A_\nu ^j(y) + A_\nu ^i(y) A_\mu^j(x)\right]\Pi _{ij}^{-kl} = \mbox{c-number}
\end{equation}
is not gauge-invariant.
\bibliographystyle{plain}

\begin{thebibliography}{99}
\bibitem{Arefeva1} I.~Y.~Aref'eva and I.~V.~Volovich, Quantum group
  chiral fields and differential Yang-Baxter equation. {\em Phys. Lett.
    B}{\bf 264}, 62 (1991)
\bibitem{Arefeva2} I.~Y.~Aref'eva and I.~V.~Volovich, Quantum group
  gauge-fields. {\em Mod. Phys. Lett. A}{\bf 6}, 893 (1991)
\bibitem{BrzezMaj:class} T.~Brzezinski and S.~Majid, Quantum group
  gauge-theory on classical spaces. {\em Phys. Lett. B} {\bf 298},
  339 (1993)
\bibitem{BrzezMaj:quant} T.~Brzezinski and S.~Majid, Quantum group
  gauge-theory on quantum spaces. {\em Commun. Math. Phys.} {\bf 157},
  591 and {\bf 167}, 235 (1995)
\bibitem{Castellani:proto} L.~Castellani, Gauge theories of quantum
  groups. {\em Phys. Lett. B}{\bf 292}, 93 (1992)
\bibitem{Castellani:UqN} L.~Castellani, $U_q(N)$ gauge theories. {\em
    Mod. Phys. Lett. A}{\bf 9}, 2835 (1993)
\bibitem{Durdevic} M.~\Dj ur\dj evi\'{c}, Quantum principal bundles and
  corresponding gauge theories. q-alg/9507021.
\bibitem{Hajac} P.~M.~Hajac, Strong connections and $U_q(2)$--Yang-Mills theory on
  quantum principal bundles. hep-th/9406129.
\bibitem{Hirayama} M.~Hirayama, Gauge field-theory of the quantum group
  $SU_q(2)$. {\em Prog. Theor. Phys.{}\bf 88}, 111 (1992)
\bibitem{IsaevPop1} A.~P.~Isaev and Z.~Popowicz, $q$-trace for quantum
  groups and $q$-deformed Yang-Mills theory. {\em Phys. Lett. B}{\bf
    281}, 271 (1992)
\bibitem{IsaevPop2} A.~P.~Isaev and Z.~Popowicz, Quantum group gauge
theories and covariant quantum algebras. {\em Phys. Lett. B}{\bf 307},
353 (1993)
\bibitem{qliealg} V.~Lyubashenko and A.~Sudbery, Quantum Lie algebras of type $A_n$.
  Preprint q-alg/9510004
\bibitem{Maillet} J.~M.~Maillet and F.~Nijhoff, Gauging the quantum
groups. {\em Phys. Lett. B}{\bf 229}, 71 (1989)
\bibitem{Watamura:BRST} S.~Watamura, Quantum deformation of BRST
  algebra. {\em Commun. Math. Phys.}{\bf 158}, 67 (1993)
\end{thebibliography}

\end{document}